\documentclass{article}
\usepackage{amssymb}
\usepackage{amsmath}

\allowdisplaybreaks

\begin{document}

\title{On the consistent interactions in $D=11$ among a graviton, a massless
gravitino and a three-form}
\author{E. M. Cioroianu\thanks{
e-mail address: manache@central.ucv.ro}, E. Diaconu\thanks{
e-mail address: ediaconu@central.ucv.ro}, S. C. Sararu\thanks{
e-mail address: scsararu@central.ucv.ro} \\
%EndAName
Faculty of Physics, University of Craiova,\\
13 Al. I. Cuza Street Craiova, 200585, Romania}
\date{}
\maketitle

\begin{abstract}
The couplings that can be introduced between a massless Rarita-Schwinger
field, a Pauli-Fierz field and an Abelian three-form gauge field in eleven
spacetime dimensions are analyzed in the context of the deformation of the
solution of the master equation.

\end{abstract}

\section{Introduction}

The D=11, N=1 SUGRA \cite{scherk}--\cite{nieuwen} has a central role
with the advent of M-theory, whose QFT (local) limit it is. It is
known that the field content of $D=11$, $N=1$ supergravity is
remarkably simple; it consists of a graviton, a massless Majorana
spin-$3/2$ field, and a three-form gauge field. The aim of this
paper is the analysis of all possible interactions in $D=11$ related
to this field content. With this purpose in mind we study first the
cross-couplings involving each pair of these sorts of fields and
then the construction of simultaneous interactions among all the
three fields. The method to be used is the deformation technique of
the solution to the classical master equation \cite{2} combined with
the local BRST cohomology \cite{5}. The requirements imposed on the
interacting theory are: space-time locality, analyticity of the
deformations in the coupling constant, Lorentz covariance,
Poincar\'{e} invariance (we do not allow explicit dependence on the
spacetime coordinates), preservation of the number of derivatives on
each field (the differential order of the deformed field equations
is preserved with respect to the free model) and the condition that
the interacting Lagrangian contains at most two space-time
derivatives (like the free one). In this paper we compute the
interaction terms to order two in the coupling constant. In this way
we obtain that the first two orders of the interacting Lagrangian
resulting from our setting
originate in the development of the full interacting Lagrangian of $D=11$, $%
N=1$ SUGRA.

\section{Construction of consistent interactions}

We begin with a free model given by a Lagrangian action, written as the sum
between the linearized Hilbert-Einstein action (also known as the
Pauli-Fierz action), the action for an Abelian three-form gauge field and
and that of a massless Rarita-Schwinger field in eleven spacetime dimensions%
\begin{eqnarray}
S_{0}^{\mathrm{L}}\left[ h_{\mu \nu },A_{\mu \nu \rho },\psi _{\mu }\right]
&=&S_{0}^{\mathrm{PF}}[h_{\mu \nu }]+S_{0}^{\mathrm{3F}}[A_{\mu \nu \rho
}]+S_{0}^{\mathrm{RS}}[\psi _{\mu }]  \notag \\
&=&\int d^{11}x\left( -\frac{1}{2}\left( \partial _{\mu }h_{\nu \rho
}\right) \left( \partial ^{\mu }h^{\nu \rho }\right) +\left( \partial _{\mu
}h^{\mu \rho }\right) \left( \partial ^{\nu }h_{\nu \rho }\right) \right.
\notag \\
&&-\left( \partial _{\mu }h\right) \left( \partial _{\nu }h^{\nu \mu
}\right) +\frac{1}{2}\left( \partial _{\mu }h\right) \left( \partial ^{\mu
}h\right)  \notag \\
&&\left. -\frac{1}{2\cdot 4!}F_{\mu \nu \rho \lambda }F^{\mu \nu \rho
\lambda }-\frac{\mathrm{i}}{2}\bar{\psi}_{\mu }\gamma ^{\mu \nu \rho
}\partial _{\nu }\psi _{\rho }\right) .  \label{fract}
\end{eqnarray}%
The theory described by action (\ref{fract}) possesses an Abelian generating
set of gauge transformations
\begin{equation}
\delta _{\epsilon ,\varepsilon }h_{\mu \nu }=\partial _{(\mu }\epsilon _{\nu
)},\qquad \delta _{\epsilon ,\varepsilon }A_{\mu \nu \rho }=\partial
_{\lbrack \mu }\varepsilon _{\nu \rho ]},\qquad \delta _{\epsilon
,\varepsilon }\psi _{\mu }=\partial _{\mu }\varepsilon ,  \label{trans}
\end{equation}%
where the gauge parameters $\left\{ \epsilon _{\mu },\varepsilon _{\mu \nu
}\right\} $ are bosonic and $\varepsilon $ is fermionic. In addition $%
\varepsilon _{\mu \nu }$\ are completely antisymmetric and $\varepsilon $ is
a Majorana spinor. The gauge algebra associated with (\ref{trans}) is
Abelian.

We observe that if in (\ref{trans}) we make the transformations $\varepsilon
_{\mu \nu }\rightarrow \varepsilon _{\mu \nu }^{\left( \theta \right)
}=\partial _{\lbrack \mu }\theta _{\nu ]}$, then the gauge variation of the
three-form identically vanishes, $\delta _{\varepsilon ^{\left( \theta
\right) }}A_{\mu \nu \rho }\equiv 0$. Moreover, if we perform the changes $%
\theta _{\mu }\rightarrow \theta _{\mu }^{\left( \phi \right) }=\partial
_{\mu }\phi $, with $\phi $\ an arbitrary scalar field, then the transformed
gauge parameters identically vanish, $\varepsilon _{\mu \nu }^{\left( \theta
^{\left( \phi \right) }\right) }\equiv 0$. Meanwhile, there is no
nonvanishing local transformation of $\phi $\ that annihilates $\theta _{\mu
}^{\left( \phi \right) }$, and hence no further local reducibility identity.
All these allow us to conclude that the generating set of gauge
transformations given in (\ref{trans}) is off-shell, second-order reducible.

In order to construct the BRST symmetry for the model under study we
introduce the fermionic ghosts $\eta _{\mu }$ and $C_{\mu \nu }$
corresponding to the gauge parameters $\epsilon _{\mu }$ and $\varepsilon
_{\mu \nu }$ respectively, the bosonic ghost $\xi $ associated with the
gauge parameter $\varepsilon $, the bosonic ghosts for ghosts $C_{\mu }$ and
the fermionic ghost for ghost for ghost $C$ due to the first- and
respectively second-order reducibility. The antifield spectrum is organized
into the antifields $\left\{ h^{\ast \mu \nu },A^{\ast \mu \nu \rho },\psi
_{\mu }^{\ast }\right\} $ of the original fields and those corresponding to
the ghosts $\left\{ \eta ^{\ast \mu },C^{\ast \mu \nu },\xi ^{\ast }\right\}
$, $C^{\ast \mu }$ and $C^{\ast }$. The antifield of the Rarita-Schwinger
field, $\psi _{\mu }^{\ast }$, is a bosonic, purely imaginary spinor.

Since both the gauge generators and the reducibility functions for this
model are field-independent, it follows that the BRST differential $s$
reduces to $s=\delta +\gamma $\ (where $\delta $ is the Koszul-Tate
differential and $\gamma $ stands for the exterior derivative along the
gauge orbits).

The action of the antifield-BRST differential $s$ can always be realized in
an anticanonical form, $s\cdot =\left( \cdot ,S\right) $, where $\left(
,\right) $ is the anticanonical structure, named antibracket, and $S$ stands
for its generator. The nilpotency of $s$\ becomes equivalent to the master
equation $\left( S,S\right) =0$. For the free model under study, $S$ reads as%
\begin{eqnarray}
S &=&S_{0}^{\mathrm{L}}+\int d^{11}x\left( h^{\ast \mu \nu }\partial _{(\mu
}\eta _{\nu )}+A^{\ast \mu \nu \rho }\partial _{\lbrack \mu }C_{\nu \rho
]}\right.  \notag \\
&&\left. +\psi ^{\ast \mu }\partial _{\mu }\xi +C^{\ast \mu \nu }\partial
_{\lbrack \mu }C_{\nu ]}+C^{\ast \mu }\partial _{\mu }C\right) .  \label{f12}
\end{eqnarray}

It has been shown in \cite{2} that if an interacting theory can be
consistently constructed, then we can associate with (\ref{f12}) a deformed
solution%
\begin{equation}
S\rightarrow \bar{S}=S+\lambda S_{1}+\lambda ^{2}S_{2}+\cdots  \label{1}
\end{equation}%
which is the BRST generator of the interacting theory (in the above $\lambda
$ is known as the coupling constant or deformation parameter)%
\begin{equation}
\left( \bar{S},\bar{S}\right) =0.  \label{2}
\end{equation}%
Projecting (\ref{1}) on the various powers in the coupling constant, we find
that the components of $\bar{S}$\ are restricted to satisfy the equivalent
tower of equations%
\begin{eqnarray}
\left( S,S\right) &=&0,  \label{3} \\
2\left( S_{1},S\right) &=&0,  \label{4} \\
2\left( S_{2},S\right) +\left( S_{1},S_{1}\right) &=&0,  \label{5} \\
&&\vdots  \notag
\end{eqnarray}%
In view of this, the construction of consistent interactions becomes
equivalent to solving equations (\ref{4})--(\ref{5}), etc. (equation (\ref{3}%
) is satisfied by hypothesis, since $S$ given by (\ref{f12}) is the solution
of the master equation for the starting free theory). The finding of
solutions to the deformation equations relies on the computation of the
local BRST cohomology of the starting free theory in ghost number zero
(ghost number is the overall degree that grades the BRST complex). According
to the decomposition given by (\ref{1}), $S_{k}$ will be called deformation
of order $k$ (of the solution to the master equation).

We have shown in \cite{PI}--\cite{PIV} that the first-order deformation $%
S_{1}$ can be decomposed as a sum of six components%
\begin{equation}
S_{1}=S_{1}^{\mathrm{h-A}}+S_{1}^{\mathrm{h-\psi }}+S_{1}^{\mathrm{A-\psi }%
}+S_{1}^{\mathrm{A}}+S_{1}^{\mathrm{\psi }}+S_{1}^{\mathrm{h}},  \label{6}
\end{equation}%
where
\begin{eqnarray}
S_{1}^{\mathrm{h-A}} &=&\int d^{11}x\left[ -kC^{\ast }\left( \partial ^{\mu
}C\right) \eta _{\mu }-\frac{k}{2}C_{\mu }^{\ast }\left( C_{\nu }\partial
^{\lbrack \mu }\eta ^{\nu ]}-\left( \partial _{\nu }C\right) h^{\mu \nu
}\right. \right.  \notag \\
&&\left. +2\left( \partial ^{\nu }C^{\mu }\right) \eta _{\nu }\right)
+kC_{\mu \nu }^{\ast }\left( h_{\rho }^{\mu }\partial ^{\rho }C^{\nu
}-\left( \partial ^{\rho }C^{\mu \nu }\right) \eta _{\rho }-\frac{1}{2}%
C_{\rho }\partial ^{\lbrack \mu }h^{\nu ]\rho }\right.  \notag \\
&&\left. +C_{\;\;\rho }^{\nu }\partial ^{\lbrack \mu }\eta ^{\rho ]}\right)
-kA_{\mu \nu \rho }^{\ast }\left( \eta _{\lambda }\partial ^{\lambda }A^{\mu
\nu \rho }+\frac{3}{2}A_{\;\;\;\lambda }^{\nu \rho }\partial ^{\lbrack \mu
}\eta ^{\lambda ]}\right.  \notag \\
&&\left. -\frac{3}{2}\left( \partial ^{\lambda }C^{\nu \rho }\right)
h_{\lambda }^{\mu }-\frac{3}{2}C^{\rho \lambda }\partial _{\left. {}\right.
}^{[\mu }h_{\lambda }^{\nu ]}\right)  \notag \\
&&\left. +\frac{k}{4}F_{\mu \nu \rho \lambda }\left( \partial ^{\mu }\left(
A^{\nu \rho \sigma }h_{\sigma }^{\lambda }\right) +\frac{k}{4!}F^{\mu \nu
\rho \lambda }h-\frac{1}{3}F^{\mu \nu \rho \sigma }h_{\sigma }^{\lambda
}\right) \right] ,  \label{7}
\end{eqnarray}

\begin{eqnarray}
S_{1}^{\mathrm{h-\psi }} &=&\int d^{11}x\left[ \bar{k}\xi ^{\ast }\left(
\partial _{\mu }\xi \right) \eta ^{\mu }-\frac{\bar{k}}{8}\left( \frac{%
\mathrm{i}}{2}\eta ^{\ast \mu }\bar{\xi}\gamma _{\mu }\xi -\xi ^{\ast
}\gamma ^{\mu \nu }\xi \partial _{\lbrack \mu }\eta _{\nu ]}\right) \right.
\notag \\
&&+\frac{\mathrm{i}\bar{k}}{4}h^{\ast \mu \nu }\bar{\xi}\gamma _{\mu }\psi
_{\nu }+\frac{\bar{k}}{8}\psi ^{\ast \mu }\gamma ^{\alpha \beta }\left( \psi
_{\mu }\partial _{\lbrack \alpha }\eta _{\beta ]}-\xi \partial _{\lbrack
\alpha }h_{\beta ]\mu }\right)  \notag \\
&&+\bar{k}\psi ^{\ast \mu }\left( \partial _{\nu }\psi _{\mu }\right) \eta
^{\nu }+\frac{\bar{k}}{2}\psi ^{\ast \mu }\psi ^{\nu }\partial _{\lbrack \mu
}\eta _{\nu ]}  \notag \\
&&-\frac{\bar{k}}{2}\psi ^{\ast \mu }\left( \partial ^{\nu }\xi \right)
h_{\mu \nu }+\frac{\mathrm{i}\bar{k}}{4}\bar{\psi}_{\mu }\gamma ^{\mu \nu
\rho }\left( \partial ^{\lambda }\psi _{\rho }\right) h_{\nu \lambda }-\frac{%
\mathrm{i}\bar{k}}{8}\bar{\psi}_{\mu }\gamma ^{\mu \nu \rho }\psi ^{\lambda
}\partial _{\lbrack \nu }h_{\rho ]\lambda }  \notag \\
&&\left. +\frac{\mathrm{i}\bar{k}}{8}\bar{\psi}^{\mu }\left( \gamma ^{\rho
}\psi ^{\nu }-\frac{\mathrm{i}\bar{k}}{2}\sigma ^{\nu \rho }\gamma _{\lambda
}\psi ^{\lambda }\right) \partial _{\lbrack \mu }h_{\nu ]\rho }-\frac{%
\mathrm{i}\bar{k}}{4}h\bar{\psi}_{\mu }\gamma ^{\mu \nu \rho }\partial _{\nu
}\psi _{\rho }\right] ,  \label{s1hpsi}
\end{eqnarray}

\begin{eqnarray}
S_{1}^{\mathrm{A-\psi }} &=&\int d^{11}x\,\left[ \frac{\tilde{k}}{2}C^{\ast
\mu \nu }\bar{\xi}\gamma _{\mu \nu }\xi -\frac{\tilde{k}}{3\cdot 4!}\psi
^{\ast \mu }F^{\nu \rho \lambda \sigma }\gamma _{\mu \nu \rho \lambda \sigma
}\xi \right.  \notag \\
&&+\frac{\tilde{k}}{9}\psi ^{\ast \mu }F_{\mu \nu \rho \lambda }\gamma ^{\nu
\rho \lambda }\xi -3\tilde{k}A^{\ast \mu \nu \rho }\bar{\xi}\gamma _{\mu \nu
}\psi _{\rho }  \notag \\
&&\left. -\frac{\tilde{k}}{4}\left( \bar{\psi}_{\mu }\gamma _{\nu \rho }\psi
_{\lambda }+\frac{1}{12}\bar{\psi}^{\alpha }\gamma _{\alpha \beta \mu \nu
\rho \lambda }\psi ^{\beta }\right) F^{\mu \nu \rho \lambda }\right] ,
\label{9}
\end{eqnarray}

\begin{equation}
S_{1}^{\mathrm{A}}=\int d^{11}x\left[ q\varepsilon ^{\mu _{1}\cdots \mu
_{11}}A_{\mu _{1}\mu _{2}\mu _{3}}F_{\mu _{4}\cdots \mu _{7}}F_{\mu
_{8}\cdots \mu _{11}}\right] ,  \label{10}
\end{equation}%
\begin{equation}
S_{1}^{\mathrm{\psi }}=\int d^{11}x\left[ m\left( \psi _{\mu }^{\ast }\gamma
^{\mu }\xi +\frac{9\mathrm{i}}{2}\psi _{\mu }\gamma ^{\mu \nu }\psi _{\nu
}\right) \right] ,  \label{11}
\end{equation}%
\begin{eqnarray}
S_{1}^{\mathrm{h}} &=&\int d^{11}x\left[ \frac{1}{2}\eta ^{\ast \mu }\eta
^{\nu }\partial _{\left[ \mu \right. }\eta _{\left. \nu \right] }+h^{\ast
\mu \rho }\left( \left( \partial _{\rho }\eta ^{\nu }\right) h_{\mu \nu
}-\eta ^{\nu }\partial _{\lbrack \mu }h_{\nu ]\rho }\right) \right.  \notag
\\
&&\left. +\mathcal{L}_{1}^{\mathrm{H-E}}-2\Lambda h\right] .  \label{expr}
\end{eqnarray}%
In formulas (\ref{7})--(\ref{11}) $k$, $\bar{k}$, $\tilde{k}$, $q$ and $m$
are some arbitrary constants. In ~(\ref{expr}) we used the notations $%
\mathcal{L}_{1}^{\mathrm{H-E}}$ and $\Lambda $ for the cubic vertex of the
Einstein-Hilbert Lagrangian and respectively for the cosmological constant.

The next equation, responsible for the second-order deformation $S_{2}$, is
precisely (\ref{5}). By direct computation we proved in \cite{PIV} that the
antibracket $\left( S_{1},S_{1}\right) $ naturally decomposes into%
\begin{eqnarray}
\left( S_{1},S_{1}\right) &=&\left( S_{1},S_{1}\right) ^{\mathrm{h-A}%
}+\left( S_{1},S_{1}\right) ^{\mathrm{h-\psi }}+\left( S_{1},S_{1}\right) ^{%
\mathrm{A-\psi }}  \notag \\
&&+\left( S_{1},S_{1}\right) ^{\mathrm{\psi }}+\left( S_{1},S_{1}\right) ^{%
\mathrm{h}}+\left( S_{1},S_{1}\right) ^{\mathrm{int}},  \label{descantip}
\end{eqnarray}%
where $\left( S_{1},S_{1}\right) ^{\mathrm{sector(s)}}$ is the projection of
$\left( S_{1},S_{1}\right) $ on the respectively mentioned \textrm{sectors(s)%
}. The consistency of the first-order deformation requires that the
constants $k$, $\bar{k}$, $\tilde{k},q$, $m$, and $\Lambda $ are subject to
the following algebraic equations:
\begin{eqnarray}
\bar{k}\left( \bar{k}-1\right)=0,\ \tilde{k}\left( k+\bar{k}\right) =0,\ m%
\tilde{k}=0,\ \tilde{k}\left( q+\frac{\tilde{k}}{3\cdot \left( 12\right) ^{3}%
}\right) =0,  \label{e2} \\
k\left( k+1\right)=0,\ \tilde{k}^{2}+\frac{\bar{k}^{2}}{32}=0,\ \tilde{k}%
^{2}-\frac{k\bar{k}}{32}=0,\ 180m^{2}-\bar{k}\Lambda =0.  \label{e1}
\end{eqnarray}%
There are two main types of nontrivial solutions to the above equations,
namely
\begin{equation}
k=-1\ \mathrm{or}\ k=0,\ \tilde{k}=\bar{k}=m=0,\ \Lambda ,q=\mathrm{arbitrary%
},
\end{equation}%
and
\begin{equation}
k=-\bar{k}=-1,\ \tilde{k}_{1,2}=\pm \frac{\mathrm{i}\sqrt{2}}{8},\ q_{1,2}=-%
\frac{4\tilde{k}_{1,2}}{\left( 12\right) ^{4}},\ m=0=\Lambda .
\label{solntr2}
\end{equation}%
The former type is less interesting from the point of view of interactions
since it maximally allows the graviton to be coupled to the $3$-form (if $%
k=-1$). For this reason in the sequel we will extensively focus on the
latter solution, ~(\ref{solntr2}), which forbids both the presence of the
cosmological term for the spin-$2$ field and the appearance of gravitini
`mass' constant. Decomposition (\ref{descantip}) allows us to write the
second-order deformation under the form%
\begin{equation}
S_{2}=S_{2}^{\mathrm{h-A}}+S_{2}^{\mathrm{h-\psi }}+S_{2}^{\mathrm{A-\psi }%
}+S_{2}^{\mathrm{\psi }}+S_{2}^{\mathrm{h}}+S_{2}^{\mathrm{int}},
\label{rf1}
\end{equation}%
with

\begin{eqnarray}
S_{2}^{\mathrm{h-A}} &=&\frac{1}{2}\int d^{11}x\left\{ \frac{1}{8}F^{\mu \nu
\rho \lambda }F_{\mu \nu \xi \pi }\left[ h_{\rho }^{\xi }h_{\lambda }^{\pi }-%
\frac{1}{3!}\delta _{\rho }^{\xi }\delta _{\lambda }^{\pi }\left( \frac{1}{4}%
h^{2}-h^{\alpha \beta }h_{\alpha \beta }\right) \right. \right.   \notag \\
&&\left. -\frac{1}{3}\delta _{\rho }^{\xi }h_{\lambda \sigma }h^{\pi \sigma }%
\right] +\frac{1}{16}F^{\mu \nu \rho \lambda }\left[ -h_{\mu \pi }h^{\xi \pi
}\left( \partial _{\nu }A_{\xi \rho \lambda }+\frac{4}{3}\partial _{\xi
}A_{\nu \rho \lambda }\right) \right.   \notag \\
&&\left. A_{\xi \rho \lambda }\partial _{\mu }\left( h_{\nu \pi }h^{\lambda
\pi }\right) +4A_{\mu \nu \xi }h_{\rho }^{\pi }\partial _{\lbrack \pi
}^{\left. {}\right. }h_{\lambda ]}^{\xi }-A_{\mu \nu \xi }h\partial
_{\lbrack \rho }^{\left. {}\right. }h_{\lambda ]}^{\xi }+2h_{\rho }^{\xi
}h_{\lambda }^{\pi }\partial _{\xi }A_{\pi \mu \nu }\right]   \notag \\
&&-\frac{1}{8}\left[ \frac{1}{3}h_{\lambda }^{\xi }h\partial _{\xi }A_{\mu
\nu \rho }+A_{\mu \xi \pi }\partial _{\nu }\left( h_{\rho }^{\xi }h_{\lambda
}^{\pi }\right) \right] F^{\mu \nu \rho \lambda }  \notag \\
&&+\frac{1}{16}\partial _{\xi }\left( h_{[\mu }^{\pi }A_{\nu \rho ]\pi
}^{\left. {}\right. }\right) \left[ \partial ^{\rho }\left( h_{\tau }^{[\xi
}A_{\left. {}\right. }^{\mu \nu ]\tau }\right) -\frac{1}{3}\partial ^{\xi
}\left( h_{\tau }^{[\mu }A_{\left. {}\right. }^{\nu \rho ]\tau }\right) %
\right]   \notag \\
&&+q_{i}\left( 3h_{\mu _{1}}^{\xi }A_{\xi \mu _{2}\mu _{3}}F_{\mu _{4}\ldots
\mu _{7}}-4h_{\mu _{1}}^{\xi }A_{\mu _{2}\mu _{3}\mu _{4}}F_{\mu _{5}\ldots
\mu _{7}\xi }\right.   \notag \\
&&\left. +\frac{1}{2}hA_{\mu _{1}\mu _{2}\mu _{3}}F_{\mu _{4}\ldots \mu
_{7}}\right) F_{\mu _{8}\ldots \mu _{11}}\varepsilon ^{\mu _{1}\ldots \mu
_{11}}  \notag \\
&&+\frac{3}{2}A^{\ast \mu \nu \rho }\left[ C_{\rho \xi }\partial _{\mu
}\left( h_{\nu \lambda }h^{\lambda \xi }\right) +\frac{3}{2}h_{\rho \xi
}h^{\lambda \xi }\partial _{\lambda }C_{\mu \nu }+2C_{\mu \lambda }h_{\nu
}^{\xi }\partial _{\lbrack \xi }^{\left. {}\right. }h_{\rho ]}^{\lambda
}\right.   \notag \\
&&-\frac{1}{2}A_{\mu \nu \lambda }\left( h^{\lambda \xi }\partial _{\lbrack
\rho }\eta _{\xi ]}+h_{\rho \xi }\partial ^{\lbrack \lambda }\eta ^{\xi
]}+2\sigma ^{\lambda \pi }\eta ^{\xi }\partial _{\lbrack \rho }h_{\pi ]\xi
}\right)   \notag \\
&&\left. \left. +A_{\mu \nu \lambda }h_{\rho \xi }\partial ^{\lbrack \lambda
}\eta ^{\xi ]}-\frac{2}{3}h^{\lambda \xi }\eta _{\xi }\partial _{\lambda
}A_{\mu \nu \rho }\right] +``more"\right\} ,  \label{s2ha}
\end{eqnarray}%
\begin{eqnarray}
S_{2}^{\mathrm{h-\psi }} &=&\int d^{11}x\left\{ \frac{\mathrm{i}}{8}\bar{\psi%
}^{\mu }\gamma ^{\sigma }\psi _{\sigma }\left[ h^{\nu \rho }\partial
_{\lbrack \mu }h_{\nu ]\rho }-h_{\rho \lbrack \mu }\partial _{\nu ]}h^{\rho
\nu }+h_{\rho \lbrack \mu }^{\left. {}\right. }\partial ^{\rho }h_{\nu
]}^{\nu }\right. \right.   \notag \\
&&\left. +\frac{1}{2}\left( \partial ^{\nu }h_{\rho \lbrack \mu }^{\left.
{}\right. }\right) h_{\nu ]}^{\rho }\right] -\frac{\mathrm{i}}{16}\bar{\psi}%
_{\mu }\gamma ^{\mu \nu \rho }\psi ^{\lambda }h\partial _{\lbrack \nu
}h_{\rho ]\lambda }  \notag \\
&&+\frac{\mathrm{i}}{64}\bar{\psi}^{\mu }\gamma _{\mu \nu \rho \lambda
\sigma }\psi ^{\nu }h^{\rho \omega }\partial ^{\lbrack \lambda }h_{\omega
}^{\sigma ]}  \notag \\
&&+\frac{\mathrm{i}}{16}\bar{\psi}^{\alpha }\gamma ^{\rho }\psi ^{\beta }%
\left[ h\left( \partial _{\lbrack \alpha }h_{\beta ]\rho }-2\sigma _{\rho
\beta }\partial _{\lbrack \alpha }h_{\lambda ]}^{\quad \lambda }\right)
-h_{\rho }^{\lambda }\partial _{\lbrack \alpha }h_{\beta ]\lambda }\right.
\notag \\
&&\left. +h_{\lambda \lbrack \alpha }\partial _{\beta ]}h_{\rho }^{\lambda
}-h_{\lambda \lbrack \alpha }\partial ^{\lambda }h_{\beta ]\rho }-\frac{1}{2}%
\left( \partial _{\rho }h_{\lambda \lbrack \alpha }^{\left. {}\right.
}\right) h_{\beta ]}^{\lambda }\right]   \notag \\
&&+\frac{\mathrm{i}}{8}\bar{\psi}_{\mu }\gamma ^{\mu \nu \rho }\left[ \left(
\partial _{\lambda }\psi _{\rho }\right) hh_{\nu }^{\lambda }+\left(
\partial _{\nu }\psi _{\rho }\right) \left( h^{\lambda \sigma }h_{\lambda
\sigma }-\frac{h^{2}}{2}\right) \right]   \notag \\
&&-\frac{\mathrm{i}}{8}\bar{\psi}_{\alpha }\gamma ^{\alpha \beta \gamma }%
\left[ h_{\beta \mu }h_{\gamma \nu }\partial ^{\mu }\psi ^{\nu }-h_{\beta
\lambda }\partial ^{\lambda }\left( h_{\gamma \sigma }\psi ^{\sigma }\right)
+\frac{3}{2}\left( \partial _{\mu }\psi _{\gamma }\right) h_{\beta \rho
}h^{\rho \mu }\right.   \notag \\
&&\left. -\frac{1}{2}\psi _{\lambda }h^{\rho \lambda }\partial _{\beta
}h_{\gamma \rho }-\frac{3}{2}\psi _{\sigma }h_{\gamma \lambda }\partial
_{\beta }h^{\lambda \sigma }\right] +\frac{\mathrm{i}}{8}h^{\ast \mu \nu
}h_{\mu }^{\rho }\bar{\xi}\gamma _{(\nu }\psi _{\rho )}  \notag \\
&&+\frac{3}{8}\psi ^{\ast \mu }\left( \partial _{\rho }\xi \right) h_{\mu
\nu }h^{\nu \rho }+\frac{1}{8}\psi ^{\ast \lbrack \mu }\psi ^{\nu ]}\left(
h_{\mu }^{\rho }\partial _{\lbrack \nu }\eta _{\rho ]}-\eta ^{\rho }\partial
_{\lbrack \mu }h_{\nu ]\rho }\right)   \notag \\
&&-\frac{1}{2}\psi ^{\ast \mu }\left( \partial _{\rho }\psi _{\mu }\right)
\eta _{\nu }h^{\nu \rho }+\frac{1}{16}\psi ^{\ast \mu }\gamma ^{\alpha \beta
}\psi _{\mu }\left( h_{\alpha }^{\rho }\partial _{\lbrack \beta }\eta _{\rho
]}-\eta ^{\rho }\partial _{\lbrack \alpha }h_{\beta ]\rho }\right)   \notag
\\
&&\left. +\frac{1}{8}\psi ^{\ast \lambda }\gamma ^{\mu \nu }\xi \left(
h_{\lambda }^{\rho }\partial _{\mu }h_{\nu \rho }-h_{\mu }^{\rho }\partial
_{\lbrack \nu }h_{\rho ]\lambda }-\frac{1}{2}h_{\mu }^{\rho }\partial
_{\lambda }h_{\nu \rho }\right) +``more"\right\} ,  \label{s2hpsi}
\end{eqnarray}%
\begin{equation}
S_{2}^{\mathrm{A-\psi }}=-\frac{\mathrm{i}}{16}\int d^{11}x\left[ 3A^{\ast
\mu \nu \rho }A_{\mu \nu }^{\quad \lambda }\bar{\xi}\gamma _{(\rho }\psi
_{\lambda )}+``more"\right] ,  \label{s2apsi}
\end{equation}%
\begin{eqnarray}
S_{2}^{\mathrm{\psi }} &=&\int d^{11}x\left\{ \frac{1}{2^{7}}\bar{\psi}%
_{\alpha }\gamma _{\rho }\psi _{\beta }\left( \bar{\psi}^{\alpha }\gamma
^{\rho }\psi ^{\beta }+2\bar{\psi}^{\alpha }\gamma ^{\beta }\psi ^{\rho }+%
\frac{1}{2}\bar{\psi}_{\mu }\gamma ^{\mu \nu \rho \alpha \beta }\psi _{\nu
}\right) \right.   \notag \\
&&-\frac{1}{2^{5}}\bar{\psi}^{\alpha }\gamma ^{\mu }\psi _{\mu }\bar{\psi}%
_{\alpha }\gamma ^{\nu }\psi _{\nu }+\frac{1}{2^{8}}\bar{\psi}_{\mu }\gamma
_{\nu \rho }\psi _{\lambda }\left( \bar{\psi}^{[\mu }\gamma ^{\nu \rho }\psi
^{\lambda ]}+\frac{1}{2}\bar{\psi}_{\alpha }\gamma ^{\alpha \beta \mu \nu
\rho \lambda }\psi _{\beta }\right)   \notag \\
&&-\frac{\mathrm{i}}{2^{5}}\psi ^{\ast \mu }\gamma ^{\alpha \beta }\left[
\psi _{\mu }\bar{\xi}\gamma _{\alpha }\psi _{\beta }-\xi \left( \bar{\psi}%
_{\mu }\gamma _{\alpha }\psi _{\beta }+\frac{1}{2}\bar{\psi}_{\alpha }\gamma
_{\mu }\psi _{\beta }\right) \right]   \notag \\
&&+\frac{\mathrm{i}}{2^{9}}\psi ^{\ast \mu }\psi ^{\nu }\bar{\xi}\gamma
_{(\mu }\psi _{\nu )}-\frac{\mathrm{i}}{3\cdot 2^{6}}\ \psi _{\lbrack \mu
}^{\ast }\gamma _{\nu \rho \lambda ]}^{\left. {}\right. }\xi \bar{\psi}^{\mu
}\gamma ^{\nu \rho }\psi ^{\lambda }  \notag \\
&&\left. +\frac{\mathrm{i}}{3\cdot 2^{7}}\psi ^{\ast \sigma }\gamma _{\mu
\nu \rho \lambda \sigma }\xi \bar{\psi}^{\mu }\gamma ^{\nu \rho }\psi
^{\lambda }+``more"\right\} ,  \label{s2psi}
\end{eqnarray}%
\begin{eqnarray}
S_{2}^{\mathrm{h}} &=&\int d^{11}x\left\{ \mathcal{L}_{2}^{\mathrm{EH}}-%
\frac{1}{4}h^{\ast \mu \nu }\left[ h_{\mu }^{\lambda }\partial _{\nu }\left(
h_{\rho \lambda }\eta ^{\lambda }\right) +\frac{1}{2}h_{\rho \lambda }\left(
\partial ^{\lambda }h_{\mu \nu }\right) \eta ^{\rho }\right. \right.   \notag
\\
&&\left. \left. +\frac{3}{2}\left( \partial _{(\mu }h_{\nu )\lambda
}-\partial _{\lambda }h_{\mu \nu }\right) h_{\rho }^{\lambda }\eta ^{\rho }%
\right] +``more"\right\} ,  \label{s2h}
\end{eqnarray}%
and the terms expressing the simultaneous interactions among all the three
types of fields amount to%
\begin{eqnarray}
S_{2}^{\mathrm{int}} &=&\int d^{11}x\left\{ \frac{\tilde{k}_{i}}{12}\left(
\bar{\psi}^{[\mu }\gamma ^{\nu \rho }\psi ^{\lambda ]}+\frac{1}{2}\bar{\psi}%
_{\alpha }\gamma ^{\alpha \beta \mu \nu \rho \lambda }\psi _{\beta }\right)
\times \right.   \notag \\
&&\times \left[ F_{\mu \nu \rho \sigma }h_{\lambda }^{\sigma }-3\partial
_{\mu }\left( h_{\nu }^{\sigma }A_{\rho \lambda \sigma }\right) \right]
\notag \\
&&-\frac{\tilde{k}_{i}}{8}hF_{\mu \nu \rho \lambda }\left( \bar{\psi}^{\mu
}\gamma ^{\nu \rho }\psi ^{\lambda }+\frac{1}{12}\bar{\psi}_{\alpha }\gamma
^{\alpha \beta \mu \nu \rho \lambda }\psi _{\beta }\right)   \notag \\
&&-\frac{\mathrm{i}\tilde{k}_{i}}{18}\psi ^{\ast \lbrack \mu }\gamma ^{\nu
\rho \lambda ]}\xi \left[ F_{\mu \nu \rho \sigma }h_{\lambda }^{\sigma
}-3\partial _{\mu }\left( h_{\nu }^{\sigma }A_{\rho \lambda \sigma }\right) %
\right]   \notag \\
&&\left. +\frac{\mathrm{i}\tilde{k}_{i}}{36}\psi _{\mu }^{\ast }\gamma ^{\mu
\nu \rho \lambda \sigma }\xi \left[ F_{\nu \rho \lambda \varepsilon
}h_{\sigma }^{\varepsilon }-3\partial _{\nu }\left( A_{\rho \lambda
\varepsilon }h_{\sigma }^{\varepsilon }\right) \right] +``more"\right\} .
\label{s2int}
\end{eqnarray}%
In formulas (\ref{s2ha})--(\ref{s2int}) $``more"$ means terms of antighost
numbers ranging from two to four.

\section{Lagrangian structure of the interacting theory}

From $\bar{S}$ (built as in (\ref{1}) on behalf of (\ref{f12}), (\ref{6})
and (\ref{rf1})) we can identify the Lagrangian gauge structure of the
interacting model. We have shown in \cite{PIV}, based on the isomorphism
between the local BRST cohomologies of the Pauli-Fierz model and
respectively of the linearized version of the vielbein formulation of
spin-two field theory and using a convenient partial gauge-fixing \cite%
{siegelfields}, that the Lagrangian action the deformed theory reads as%
\begin{eqnarray}
S^{\mathrm{L}} &=&\int d^{11}x\left[ \frac{2}{\lambda ^{2}}eR\left( \Omega
\left( e\right) \right) -\frac{\mathrm{i}e}{2}\bar{\psi}_{\mu }\Gamma ^{\mu
\nu \rho }D_{\nu }\left( \frac{\Omega +\hat{\Omega}}{2}\right) \psi _{\rho }-%
\frac{e}{48}\bar{F}_{\mu \nu \rho \lambda }\bar{F}^{\mu \nu \rho \lambda
}\right.  \notag \\
&&-\frac{\lambda \tilde{k}_{i}}{96}e\left( \bar{F}_{\mu \nu \rho \lambda }+%
\hat{F}_{\mu \nu \rho \lambda }\right) \left( \bar{\psi}_{\alpha }\Gamma
^{\alpha \beta \mu \nu \rho \lambda }\psi _{\beta }+2\bar{\psi}^{[\mu
}\Gamma ^{\nu \rho }\psi ^{\lambda ]}\right)  \notag \\
&&\left. -\frac{4\lambda \tilde{k}_{i}}{\left( 12\right) ^{4}}\varepsilon
^{\mu _{1}\mu _{2}\cdots \mu _{11}}\bar{A}_{\mu _{1}\mu _{2}\mu _{3}}\bar{F}%
_{\mu _{4}\cdots \mu _{7}}\bar{F}_{\mu _{8}\cdots \mu _{11}}\right]
\label{lagfull}
\end{eqnarray}

and, moreover, it is invariant under the gauge transformations%
\begin{equation}
\frac{1}{\lambda }\bar{\delta}_{\epsilon ,\varepsilon }e_{\;\;\mu }^{a}=\bar{%
\epsilon}^{\rho }\partial _{\rho }e_{\;\;\mu }^{a}+e_{\;\;\rho }^{a}\partial
_{\mu }\bar{\epsilon}^{\rho }+\epsilon _{\;\;b}^{a}e_{\;\;\mu }^{b}+\frac{%
\mathrm{i}\lambda }{8}\bar{\varepsilon}\gamma ^{a}\psi _{\mu },  \label{id6a}
\end{equation}%
\begin{equation}
\bar{\delta}_{\epsilon ,\varepsilon }\bar{A}_{\mu \nu \rho }=\partial
_{\lbrack \mu }\bar{\varepsilon}_{\nu \rho ]}+\lambda \left[ \bar{\epsilon}%
^{\lambda }\partial _{\lambda }\bar{A}_{\mu \nu \rho }+A_{\lambda \lbrack
\mu \nu }\left( \partial _{\rho ]}\bar{\epsilon}^{\lambda }\right) -\tilde{k}%
_{i}\bar{\varepsilon}\Gamma _{\lbrack \mu \nu }\psi _{\rho ]}\right] ,
\label{tr3fcurb}
\end{equation}%
\begin{eqnarray}
\bar{\delta}_{\epsilon ,\varepsilon }\psi _{\mu } &=&D_{\mu }\left( \hat{%
\Omega}\right) \varepsilon +\lambda \left[ \left( \partial _{\rho }\psi
_{\mu }\right) \bar{\epsilon}^{\rho }+\psi _{\rho }\partial _{\mu }\bar{%
\epsilon}^{\rho }+\frac{1}{4}\gamma ^{ab}\psi _{\mu }\epsilon _{ab}\right.
\notag \\
&&\left. +\frac{\mathrm{i}\tilde{k}_{i}}{9}\Gamma ^{\nu \rho \lambda
}\varepsilon \hat{F}_{\mu \nu \rho \lambda }-\frac{\mathrm{i}\tilde{k}_{i}}{%
72}\Gamma _{\mu \nu \rho \lambda \sigma }\varepsilon \hat{F}^{\nu \rho
\lambda \sigma }\right] .  \label{trfullmajcurb}
\end{eqnarray}%
The deformed gauge transformations remain second-order reducible. The entire
gauge structure of interacting theory can be found in \cite{PIV}.

It is clear now that the interacting model resulting from our
cohomological approach is nothing but $D=11$, $N=1$ SUGRA
\cite{scherk}--\cite{nieuwen}.

\section{Conclusion}

In this paper we have shortly presented the problem of the cohomological
BRST approach to the construction of consistent interactions in eleven
spacetime dimensions that can be added to a free theory describing a
massless spin-$2$ field, a massless (Rarita-Schwinger) spin-$3/2$ field, and
an Abelian $3$-form gauge field. The couplings are obtained under the
hypotheses of analyticity in the coupling constant, space-time locality,
Lorentz covariance, Poincar\'{e} invariance, and the derivative order
assumption (the maximum derivative order of the interacting Lagrangian is
equal to two, with the precaution that each interacting field equation
contains at most one spacetime derivative acting on gravitini). Our main
result is that if we decompose the metric like $g_{\mu \nu }=\sigma _{\mu
\nu }+\lambda h_{\mu \nu }$, then we can indeed couple the $3$-form and the
gravitini to $h_{\mu \nu }$ in the space of formal series with the maximum
derivative order equal to two in $h_{\mu \nu }$ such that the resulting
interactions agree with the well-known $D=11$, $N=1$ SUGRA couplings in the
vielbein formulation.

\section*{Acknowledgments}

This work is partially supported by the European Commission FP6 program
MRTN-CT-2004-005104 and by the grant AT24/2005 with the Romanian National
Council for Academic Scientific Research (C.N.C.S.I.S.) and the Romanian
Ministry of Education and Research (M.E.C.).

\end{document}